\documentstyle [12pt,epsf]{article}
\newcommand{\ga}{\alpha}
\newcommand{\gd}{\delta}
\newcommand{\gD}{\Delta}
\newcommand{\gep}{\epsilon}

\newcommand{\gi}{\imath}
\newcommand{\gl}{\lambda}
\newcommand {\gr} {\rho}

\newcommand {\mC} {{\bf C}}
\newcommand {\mD} {{\bf D}}
\newcommand {\mH} {{\bf H}}
\newcommand {\mI} {{\bf I}}
\newcommand {\mM} {{\bf M}}
\newcommand {\mQ} {{\bf Q}}
\newcommand {\Tr} {\mbox{Tr}}
\newcommand {\Var} {\rm{Var}}
\begin{document}
\begin{titlepage}
\begin{flushright}
       {\bf UK/97-12}  \\
 June 1997   \\
       hep-lat/9707001\\
\end{flushright}
\begin{center}
 
{\bf {\LARGE Pad\'e - Z$_2$ Estimator of Determinants}}
 
\vspace{1cm}
 
{\bf C. Thron, S.J. Dong, K. F. Liu, and H. P. Ying \footnote
{On leave from: Zhejiang Institute of Modern Physics,
 Zhejiang University, China}} \\[0.5em]
 {\it  Dept. of Physics and Astronomy  \\
  Univ. of Kentucky, Lexington, KY 40506}
 
\end{center}
 
\vspace{0.4cm}
 
\begin{abstract}
 
We introduce the Pad\'e--Z$_2$ (PZ)
stochastic estimator for calculating determinants and determinant ratios.
The estimator is applied to the calculation of fermion determinants from the
two ends of the Hybrid Monte Carlo trajectories with pseudofermions.
Our results on the $8^3 \times 12$ lattice with Wilson action show that
the statistical errors from the stochastic estimator can be
reduced by more than an order of magnitude by employing an unbiased
variational subtraction scheme which utilizes the off-diagonal matrices
from the hopping expansion.
Having been able to reduce the error of the determinant ratios to
about 20 \% with a relatively small number of noise vectors,
this may become a feasible algorithm for simulating dynamical
fermions in full QCD.
We also discuss the application to the density of states in
Hamiltonian systems.
\end{abstract}
 
\vfill
\end{titlepage}
 
\section{Introduction}
At present, lattice gauge Monte Carlo calculation is still the
only
viable and practical means of solving QCD and computing hadron masses and
matrix elements non-perturbatively. As such, there is a perpetual need of
sharpening the tools to tackle the numerically intensive aspects of the
computation, especially those pertaining to dynamical fermions.
 
The Euclidean functional integration formulation of the quantum field
theory of gauge bosons and fermions has the generic partition
function
 
\begin{eqnarray}
 ~ Z = \int [dU] [d\bar{\psi}] [d\psi] e^{-S_g - \bar{\psi} \mM \psi}.
 ~~~~~~~~~~~~~~~~~
\end{eqnarray}
where U is the gauge link variable and $S_g$ the gauge part of the
action. Given that the fermion part of the action is quadratic in the
Grassmann numbers $\bar{\psi}$ and $\psi$, they can be formally
integrated out to give a fermion determinant, i.e.
\begin{equation} \label{par_det}
~ Z = \int [dU] det \mM[U] e^{-S_g}.
\end{equation}
Since numerically the computation of the fermion determinant is
much more demanding a task than the updating of gauge links U,
it is often approximated by a constant.
This is known as the {\it quenched approximation} which has previously
been interpreted as tantamount to neglecting the internal quark loops.
Recently, Sexton and Weingarten~\cite{sw97} have advanced the view
that it actually corresponds to the inclusion of the leading terms in
the loop expansion which are commensurate with the size of loops in the
gauge action. This leads to a shift in $\beta$ or the bare coupling
constant. Although a number of low-energy quantities, such as
hadron masses~\cite{bcs93}, weak matrix elements~\cite{fly97,sha97}, and
hadron structure ~\cite{ldd94,dll95} are reproduced reasonably well
(within 6\% to 15\% in many cases) in the
quenched approximation, we know that the chiral log behaviors of these
quantities~\cite{sha90,bg92}, the $\eta'$ mass, and the phase transition
~\cite{uka97} depend crucially on the full inclusion of dynamical
fermions. Thus, solving full QCD with dynamical quarks remains
a desirable and challenging ultimate goal.
 
In view of the perceived difficulty of calculating determinants accurately,
all the existing working algorithms have avoided calculating them directly.
Instead, pseudofermions~\cite{wp81} and local bosons~\cite{lus94} are
introduced to bonsonize the determinantal effects.
For example, the current state of the art
algorithm -- Hybrid Monte Carlo (HMC)~\cite{dkp87} transforms
the partition function in Eq. (\ref{par_det}) to the
following one for 2 degenerate flavors
\begin{eqnarray}
 ~ Z = \int [dU] [d\phi] [d\phi^{\star}] e^{-S_G -\phi^{\star}
     (\mM^\dagger \mM)^{-1}  \phi}~.~~~~~~~~~~~~~~~~~~~~~~~~~~~~~~~~~~~
\end{eqnarray}
where $\phi$ is a pseudofermion variable which can be generated using
a Gaussian heatbath. This gives rise to the pseudofermion force which
acts over the course of
molecular dynamics trajectories to update the gauge field and the fermion
matrix $\mM$ which in turn enables the updating of the $\phi$ field.
 
 
On the other hand, it may be desirable to admit the determinantal effects
directly without resorting to the superfluous degrees of freedom from
pseudofermions. This can be done in principle with the partition function
in Eq. (\ref{par_det}) rewritten as
\begin{eqnarray}
 ~ Z &=& \int [dU] e^{-S_G + \Tr \log \mM},~~~~~~~
\label{new_par_det}
\end{eqnarray}
and $\Tr \log \mM$ which reflects the dynamical fermions is taken as an
additional part of the gauge action. It is shown in~\cite{ldt97} that a
HMC-like algorithm based on the partition function in Eq. (\ref{new_par_det})
is valid provided that $\Tr \log\mM$ can be
estimated without bias. However, the task at first sight
appears daunting. First of all, one needs an efficient algorithm to
calculate $\Tr \log\mM$. This is apparently much more intensive numerically
than calculating $(M^{\dagger}M)^{-1}\phi$ in the pseudofermion approach.
In addition, the demands on the accuracy of $\Tr \log\mM$ are very stringent.
Since the relative error of $\det \mM$ is the absolute error in
$\Tr \log \mM$, a 20\% error in $\det \mM$ for example
would require calculating $\Tr \log \mM$
(which is  of the order N) to within 0.2.
Luckily, the Monte Carlo updating involves only determinant ratios, and not
the determinants themselves. One would expect that an accuracy of $\sim$ 0.2
should be somewhat easier to achieve for the difference
$\Tr \log \mM_1 - \Tr \log \mM_2$ than for each term separately.
 
 
We shall present in this manuscript an efficient stochastic algorithm to
estimate $\Tr \log \mM$ which has the potential of achieving the kind of
accuracy ($\sim$ 0.2) in $\Tr \log$ difference with relatively small
$(\sim 500)$ number of noise vectors.
This new algorithm invokes the Pad\'e approximation for the $\log\mM$
and uses complex Z$_2$ noise to estimate the trace, as introduced in Sec. 2.
We have tested it by calculating the determinants and determinant ratios of
fermion
matrices from both ends of randomly chosen molecular dynamics trajectories
generated by
Hybrid Monte Carlo with pseudofermions. We also applied the method to
Wilson fermions
on an  $8^3 \times 12$ lattice, and studied its dependence on the rank of
the Pad\'e
expansion and the number of noise vectors. In Sec. 3 we introduce
an unbiased variational subtraction scheme which is based on the
subtraction of traceless terms in the hopping parameter expansion. We
find that
this can reduce the statistical error by an order of magnitude leading to
an error in the range of 0.2 -- 0.3 with 400 -- 600 noise vectors.
These results are presented in Sec. 4. We should mention that there exist
other stochastic estimators for determinants. Ref.~\cite{sw97} uses the
Chebyshev polynominal to expand $\log \mM^{\dagger}\mM$ and gaussian
noise to estimate the trace. Ref.~\cite{bfg96} uses  Z$_2$ noise
and the Riemann-Stieltjes integral to estimate the $\Tr \log \mM$. The
subtraction scheme we introduce here is applicable to both of these
approaches. A discussion of application to density of states in
Hermitian Hamiltonian systems is presented in Sec. 5. Sec. 6 gives the
conclusions and outlook.

\vspace{0.05in}
\section{The basic Pad\'e -- Z$_2$ Method }
\subsection{Pad\'e approximation}
The starting point for the current algorithm is the
Pad\'e approximation of the logarithm function.
The Pad\'e approximant to $\log (z)$  of order $[K,K]$ at $z_0$
is a rational function $N(z)/D(z)$ where deg~$N(z)$~=~deg~$D(z)$ = $K$,
whose value and first $2K$ derivatives agree with $\log z$ at the
specified point $z_0$. When the Pad\'e approximant $N(z)/D(z)$
is expressed in partial fractions, we obtain
\begin{equation}
\log z \approx b_0 + \sum_{k=1}^{K}\left( \frac{b_k}{z +c_k} \right),
\label{Pade_apprx}
\end{equation}
whence it follows
\begin{equation}    \label{Tr_apprx}
\mbox {log det }\mM =
\mbox {Tr log}\mM \approx b_0 \Tr {\bf I} +
               \sum_{k=1}^K  b_k\cdot \Tr (\mM +c_k{\bf I})^{-1}.
\end{equation}
 
For the purpose of Monte Carlo updating, only the ratio of determinants
is needed. In this case, the log of the determinant ratio
det$\mM_1$/det$\mM_2$ is approximated as,
\begin{eqnarray}
\log \{ \mbox {det} \mM_1/\mbox {det} \mM_2\}
  & = &\mbox {Tr}[\log \mM_1 -\log \mM_2] \nonumber\\
&\approx& \sum_{k} b_k \left(
    \Tr(\mM_1 +c_k{\bf I})^{-1} - \Tr (\mM_2 +c_k{\bf I})^{-1} \right),
    \label{Pade_expand}
\end{eqnarray}
where $\mM_1$ and $\mM_2$ are matrices at the beginning and end of an HMC
trajectory for example.
 
This approximation is accurate so long as the eigenvalues
of the matrices $\mM_1$ and $\mM_2$ all lie in the region in the complex
plane where the Pad\'e approximation is accurate.  If we define $\gep(z)$
to be the difference between the right and left-hand sides of
Eq. (\ref{Pade_apprx}), then the
error in the approximation in Eq. (\ref{Tr_apprx}) is
\begin{equation}
\gD_{\Tr \log \mM} = \sum_n \gep(\gl_n),
\end{equation}
where $\{ \gl_n \}$
are the eigenvalues of  $\mM$.
 
\begin{figure}[h]
\setlength\epsfxsize{120mm}
\epsfbox{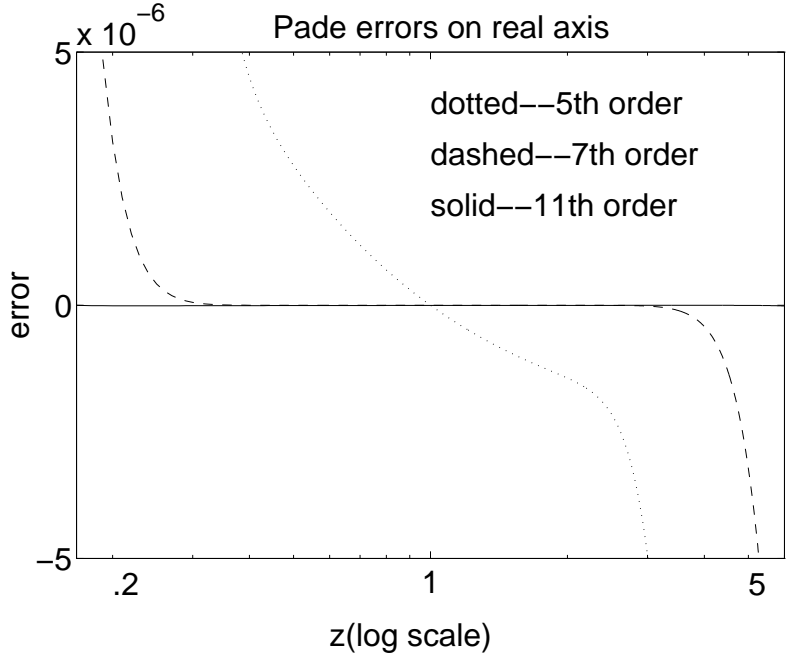}
\caption{$\epsilon(z)$ for the Pad\'e approximation of $\log z$
on the positive real axis for different orders of the Pad\'e expansion at
$z_0 = 1.0$.}
\end{figure}
 
\newpage
\clearpage
The accuracy of the Pad\'e approximation is graphically illustrated in
Figures 1-6.  In Fig. 1, we plot $\gep(z)$ for different orders of the
Pad\'e
approximation on the positive real axis. We see that $\gep(z)$ for the
5th order Pad\'e reaches quickly to the order of $10^{-6}$ around the expansion
point $z_0 = 1$, whereas that of the 7th order does not reach $10^{-6}$ until
z is smaller than 0.3 and greater than 3. For the 11th order, the domain
for which $\gep(z) < 10^{-6}$ is extended to between 0.1 and 10.
 
\begin{figure}[h]
\setlength\epsfxsize{120mm}
\epsfbox{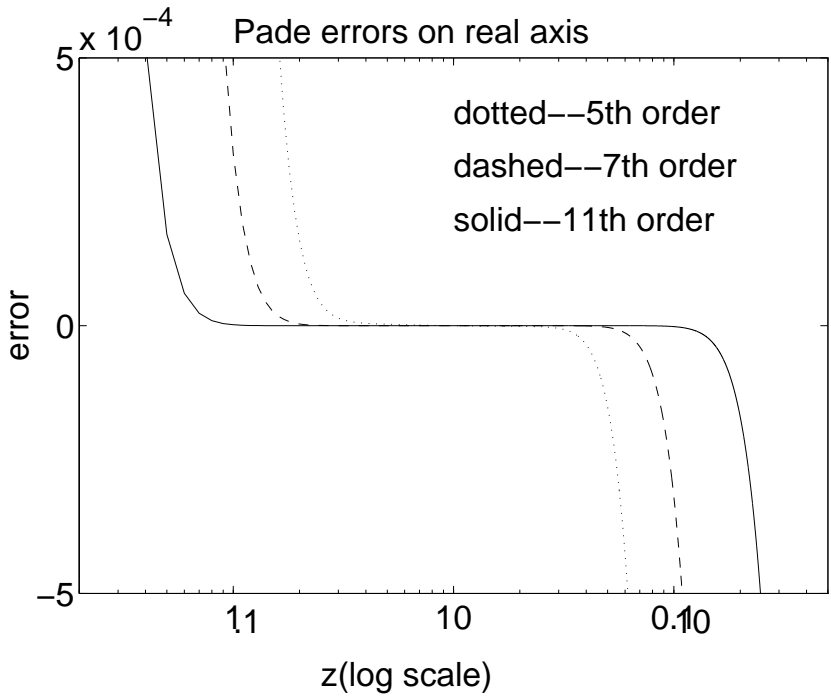}
\caption{The same as Fig. 1 with a larger domain of $z$.}
\end{figure}
 
 Fig. 2 shows
the same plot with a larger domain of z where $\gep(z)$ is larger when it
is farther away from the expansion point.
Note that in these figures $\gep(1/z) =
-\gep(z)$, i.e. the error function
is antisymmetric under the transformation $z \rightarrow 1/z$ (due to the
corresponding antisymmetry of the logarithm).
Hence, if the expansion point is suitably chosen near the `center' of the
eigenvalue distribution, we shall get error cancellation.
The coefficients of the Pad\'e
approximation $b_k$ and $c_k$ in Eq. (\ref{Pade_apprx}) for 5th, 7th, 9th,
and 11th orders used to produce Figs. 1 and 2 are tabulated in Table 1.

   In Figs. 1 and 2, the Pad\'e expansion point
is chosen as $z_0 = 1$.  The error functions $\gep_{z_0}(z)$  for other
expansion points
are identical, modulo a change of scale.  This is
due to the fact that $\log z = \log (z/z_{0}) + \log(z_{0})$,
and hence the Pad\'e  approximation $P_{z_0}(z)$ for $\log z$ around $z_0$
is equal to
\begin{equation}
 P_{z_0}(z) = P_1(z/z_{0}) + \log(z_{0}).
\label{z0_change}
\end{equation}
 It follows that
\begin{equation}
\gep_{z_0}(z) = P_1(z/z_0) - \log(z/z_0) = \gep(z/z_0).
\end{equation}

 We see from these figures that as long as one has some notion about
the domain of the eigenvalues
of the matrix $\mM$ and the desirable level of accuracy one needs for
 $\Tr \log \mM$, one can decide on an appropriate order of the Pad\'e
approximation and the expansion point $z_0$.
\newpage
\clearpage
\begin{table} 
\caption{The coefficients $b_k$ and $c_k$ of Pad\'e expansions $P[K,K]$
for log$z$ at $z_0 = 1.0$.}

\begin{tabular}{ c c c|c c } \hline\hline
 $k$ &  $b_k$  &  $c_k$  &  $b_k$  &  $c_k$  \\ \hline
 ~ & $P[5,5]$ & ~ & $P[7,7]$ ~~~~~ & \\
 ~ & ($b_0=4.566666667$)   & ~ & ($b_0=5.185714286$) \\ \hline
  1  & 0.130408172495391  & 0.0492189449629343
                          & 0.06816753269 & 0.02611045152 \\
  2  & 0.404437841802115  & 0.299993435237384
                          & 0.1844456100 & 0.1484146919  \\
  3  & 1.13777777850000   & 1.00000000000000
                          & 0.3863893344 & 0.4226317870  \\
  4  & 4.49394268525114   & 3.33340630084376
                          & 0.8359183641 & 1.0000000000  \\
  5  & 53.8334305460671   & 20.3173813294693
                          & 2.163220583 & 2.3661258591  \\
  6  & ~ & ~
                          & 8.373644726 & 6.737877409   \\
  7  & ~ & ~
                          & 99.98821380 & 38.29883980  \\ \hline
  ~ & $P[9,9]$ & ~ & $P[11,11]$ ~~~  & \\
  ~ & ($b_0=5.65793650793651$) & ~ & ($b_0=6.03975468975469$) \\  \hline
  1  & .041962745788105575 & .01617742288114674
                            & .02845031368729848 & .01100547288317344 \\
  2  & .10717746225992    & .08930616263474539
                            & .07053085455625973 & .05984825317707202  \\
  3  & .20024123115789    & .2396401470008922
                            & .1244662250449701 & .1559677956367871    \\
  4  & .35622568741554    & .5102849384064862
                            & .2021047007456418 & .3165723758668845    \\
  5  & .6604787100025195  & 1.0000000000000000
                            & .3261128629248666 & .5753698412085129   \\
  6  & 1.36804295333940783 & 1.959689429836578
                            & .5458501735606061 & 1.000000000000000   \\
  7  & 3.48685872889035930 & 4.172923495979481
                            & .9850850802286254 & 1.738012541463052    \\
  8  & 13.4381848942707415 & 11.9743554641257
                            & 2.016649317776546 & 3.158835312972759    \\
  9  & 160.340827693254954 & 61.81454285684810
                            & 5.116602053912441 & 6.411580005456822   \\
  10 & ~ & ~
                            & 19.69138119799427 & 16.70892543916556    \\
  11 & ~ & ~
                            & 234.8927672230579 & 90.86388296216937   \\
\hline \hline
\end{tabular}
\end{table}
 
    The Pad\'e approximation of the logarithm is not limited to the real
axis. It applies
equally well to the complex plane, except near the branch cut.
Pad\'e approximation of the
logarithm about a positive real $z_0$ corresponds to a branch
cut along the negative real axis, and the poles of the
Pad\'e functions all lie on the negative real axis. Using coefficients
from Table 1 which are obtained from expansion about $z_0 = 1$, we plot
the Pad\'e approximated $\log z$ along the unit circle as a function
of arg(z). The real part as shown in Fig. 3 reveals the pole at $z = -1$
for all the Pad\'e functions with different orders.
Fig. 4 shows the
imaginary part. The imaginary part of the $\log$ function has a
discontinuity from $-\pi$ to $\pi$ across the negative real axis,
while the Pad\'e approximations
have discrete poles. Thus, the Pad\'e approximation fails near the
branch cut. Nevertheless, it works for cases away from the branch cut.
In some
cases, it maybe desirable to place the branch cut in a different
location, say along the ray arg($z$)=$\theta$.  This is easily
done by choosing the expansion point $z_0$ as $e^{-i\theta}$, as
can be seen from Eq. (\ref{z0_change}). As long as there are no
eigenvalues along the ray arg($z$)=$\theta$ for the matrix $\mM$, one
can apply the Pad\'e approximation outlined above to calculate determinants
which are negative or complex.

It is worth mentioning that the Pad\'e approximation furnishes a
much more accurate
global approximation to the logarithm than that obtained from
the ``Green function''  method~\cite{wbd80}.
\newpage
\clearpage
\begin{figure}[h]
\setlength\epsfxsize{120mm}
\epsfbox{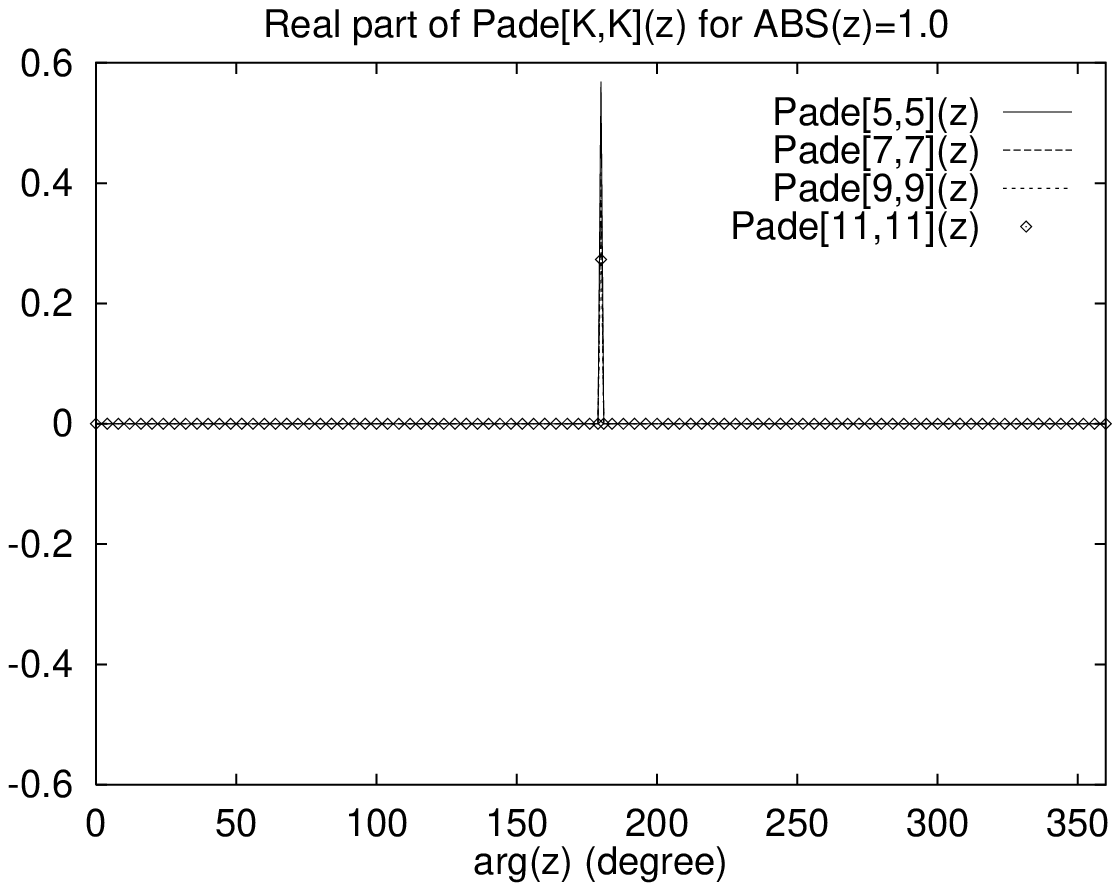}
\caption{The real part of Pad\'e approximated $\log z$ along a circle of
radius 1 as a function of arg(z) for several different orders of the
Pad\'s expansion. The expansion point is at $z_0 = 1$.}
\epsfbox{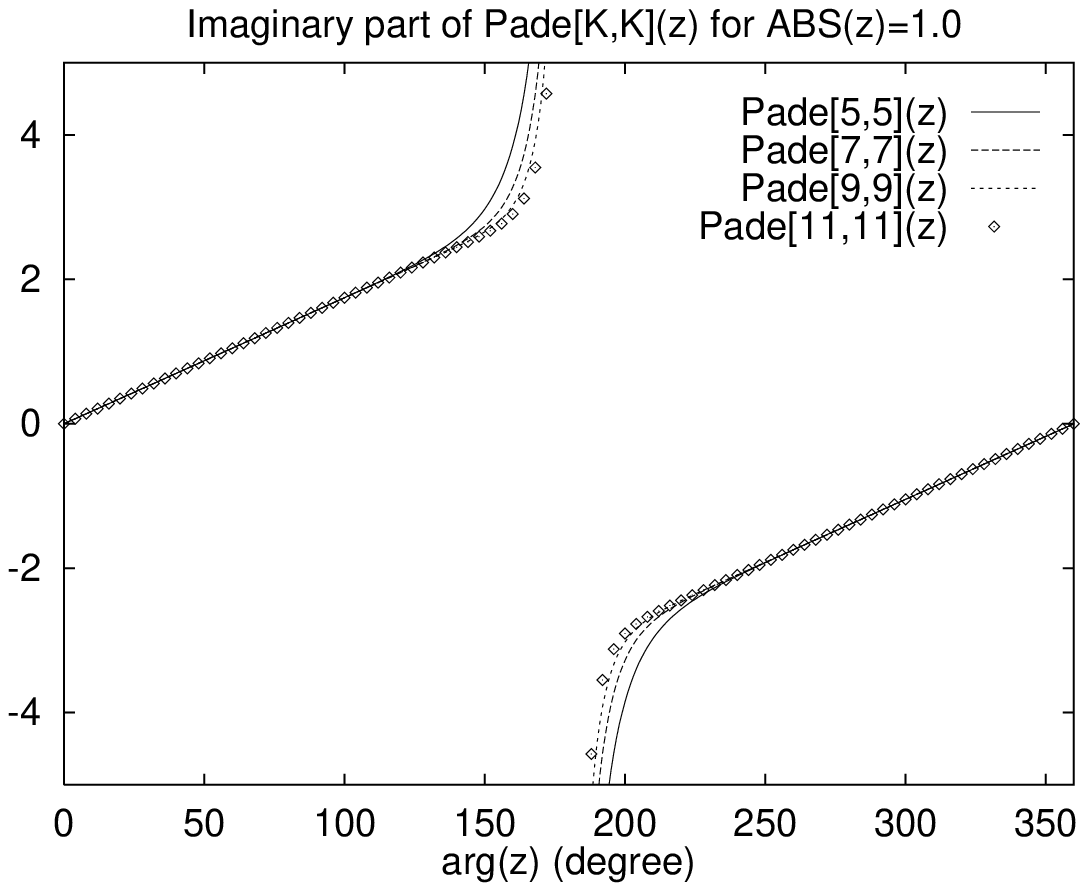}
\caption{The imaginary part of Pad\'e approximated $\log z$ along a circle of
radius 1 as a function of arg(z) for several different orders of the
Pad\'s expansion.The expansion point is at $z_0 = 1$.}
\end{figure}
\newpage
\clearpage
\begin{figure}[th]
\setlength\epsfxsize{120mm}
\epsfbox{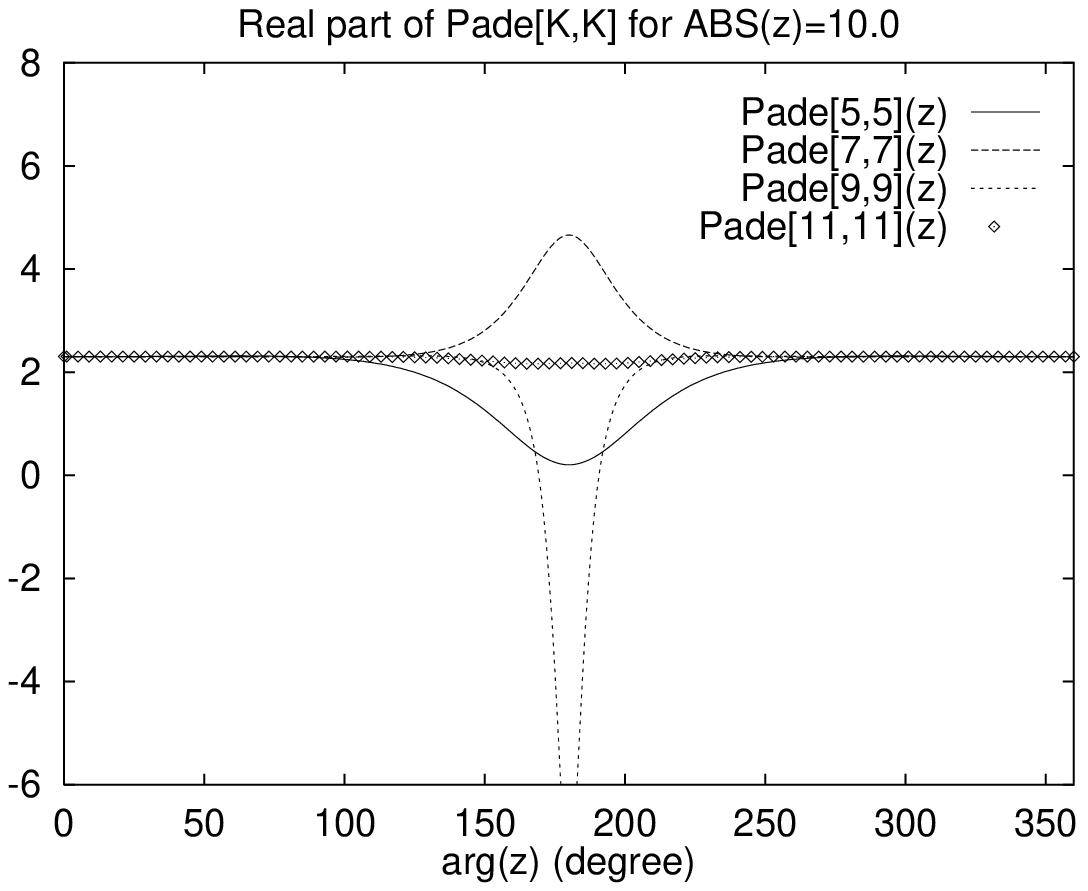}
\caption{The same as in Fig. 3 for the circle of radius 10.}
\epsfbox{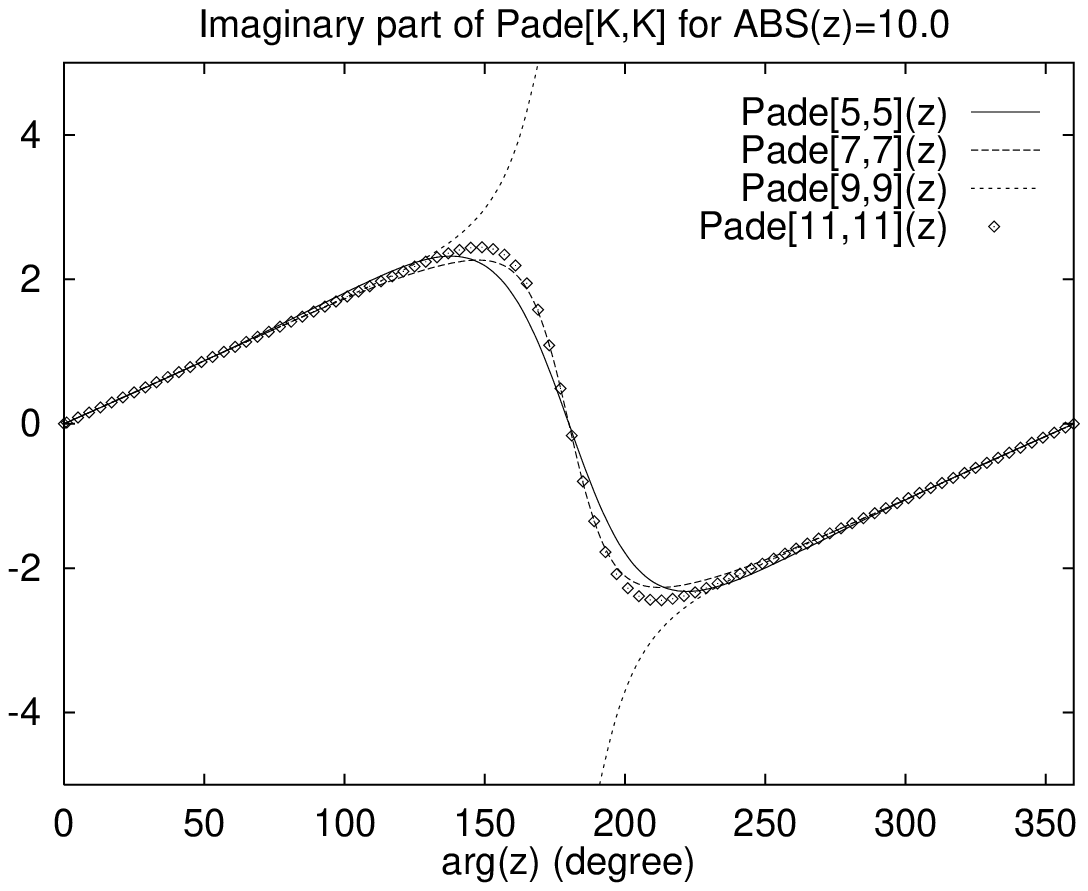}
\caption{The same as in Fig. 4 for the circle of radius 10.}
\end{figure}
\newpage
\clearpage
\begin{figure}[th]
\setlength\epsfxsize{120mm}
\epsfbox{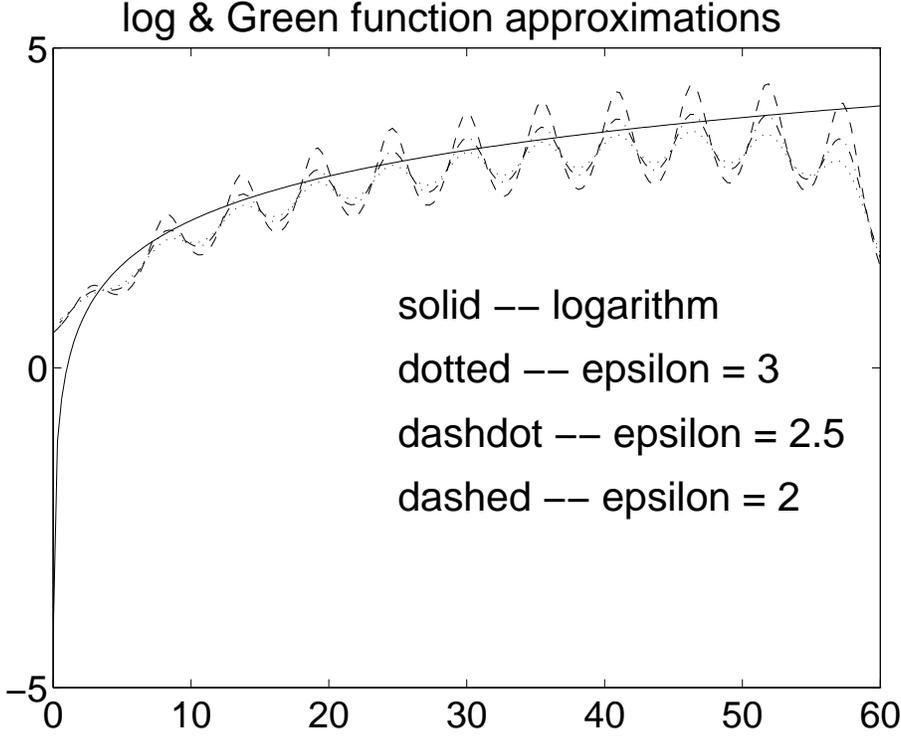}
\caption{The Green function approximation to $\log$ with 11 terms and
several $\gep$ are compared to the $\log$ function itself.}
\end{figure}

The Green
function
approximation makes use of the fact that
\begin{equation}
\frac{1}{\pi}\lim_{\gep \rightarrow 0}
                 \frac{1}{\gl - \gl_0 - \gi \gep} = \gd(\gl - \gl_0),
\end{equation}
whence it follows
\begin{eqnarray}
\mbox {Tr log}\mM &=&
   \int_{\gl_{min}}^{\gl_{max}} \log \gl \cdot \gr (\gl) d \gl \\
&\approx&
   \sum_{i} \log \gl_i \cdot \frac{\gD \gl}{\pi} \mbox{Im}
   \int_{\gl_{min}}^{\gl_{max}}
        \frac{\gr(\gl)}{\gl - \gl_i - \gi \gep}d\gl \\
&=&
   \frac{1}{\pi} \sum_{i} \log \gl_i \cdot \gD \gl~\mbox{Im}
   \left[ \Tr (\mM - \gl_i - \gi \gep)^{-1} \right],
\end{eqnarray}
where $\gl_i$ are evenly spaced on $[\gl_{min}, \gl_{max}]$, and $\gep$
is
a small parameter.
 
This approximation may be rewritten as
\begin{equation}
\int_{\gl_{min}}^{\gl_{max}} \log \gl \cdot \gr (\gl) d \gl
   \approx \int_{\gl_{min}}^{\gl_{max}} f_{\gep}(\gl) \gr (\gl) d \gl,
\end{equation}
where
\begin{equation}
f_{\gep}(\gl) = \frac{\gD \gl}{\pi} \sum_i \frac{\gep \log \gl_i}{(\gl -
\gl_i)^2 + \gep^2}.
\end{equation}
 
Figure 7 shows that the function $f_{\gep}(\gl)$
with eleven terms and several $\epsilon$ furnishes a
very poor approximation to the logarithm on the interval
$1/60 < \gl < 60 $.  By contrast,
Figure 2 shows that an
eleven-term Pad\'e expansion on the same interval
approximates the logarithm to  $\pm .0005$ at worst,
and to much higher accuracy on most of the interval.
Also, the Green function method is only
applicable if the eigenvalues of the matrix are {\em real}, while the
Pad\'e approximation also holds for matrices with {\em complex}
eigenvalues.
 
  \subsection{Complex Z$_2$ noise trace estimation}
Exact computation of the trace inverse for $N \times N$ matrices is very
time consuming  for matrices
of size $N \sim 10^6$.  However, the complex Z$_2$ noise method
has been shown to provide an efficient stochastic estimation of the
trace~\cite{dl94,dll95,Eick96}. In fact, it has been proved to be an
optimal choice for the noise, producing a {\it minimum}
variance~\cite{bmt94,hut89}.
 
The complex Z$_2$ noise estimator can be briefly described as follows
\cite{dl94,bmt94}.
We construct L noise vectors
$\eta^1,\eta^2,\cdots, \eta^L$ where
$\eta^j = \{ \eta^j_1, \eta^j_2, \eta^j_3, \cdots, \eta^j_N \}^{T}$,
as follows. Each element $\eta^j_n$ takes one of the four values
$\{\pm 1 ,\,\pm \imath \}$  chosen independently with equal probability.
It follows from the statistics of $\eta^j_n$ that
\begin{eqnarray}
E[<\eta_n>] \equiv E[\frac {1}{L} \sum_{j=1}^L \eta^j_n]=0,~~~~~~~~
E[<\eta_m^{\star} \eta_n>] \equiv
        E[\frac {1}{L} \sum_{j=1}^L \eta_m^{\star
j}\eta_n^j]=\delta_{mn}.~~~~~~~~~~~~~
\end{eqnarray}
The vectors can be used to construct an unbiased estimator for the trace
inverse of a given matrix $M$  as follows:
\begin{eqnarray*}
E[<\eta^\dagger  \mM^{-1} \eta>]&\equiv&
E[\frac 1L \sum_{j=1}^L \sum_{m,n=1}^N \eta^{\star j}_m M^{-1}_{m,n}
\eta^j_n]
~~~~~~~~~~~~~~~~~~~~~~~~~~~~~~~~~~~~~~~~~~~~~~~\\
&=&\frac 1L \sum_{j=1}^L (\sum_{n=1}^N M^{-1}_{n,n}) E[\eta^{\star j}_n
\eta^j_n] +
\frac 1L \sum_j^L (\sum_{m\neq n}^N M^{-1}_{m,n}) E[\eta^{\star j}_m
\eta^j_n ]\\
&=& \sum_n^N M^{-1}_{n,n}  +
(\sum_{m\neq n}^N M^{-1}_{m,n})[\frac 1L \sum_j^L \eta^{\star j}_m
\eta^j_n]\\
=&& \mbox {Tr }\mM^{-1}.
\end{eqnarray*}
The variance of the estimator is shown to be \cite{bmt94}
\begin{eqnarray*}
\sigma^2_M &\equiv& \Var[<\eta^\dagger \mM^{-1}\eta>]=
E\left[ |<\eta^\dagger \mM^{-1}\eta>-\mbox {Tr }\mM^{-1}|^2\right]
~~~~~~~~~~~~~~~~~~~~~~~~~~~~\\
 &=& \frac 1L \sum_{m\neq n}^N M^{-1}_{m,n}(M^{-1}_{m,n})^\star
      = \frac 1L \sum_{m\neq n}^N |M^{-1}_{m,n}|^2~.
\end{eqnarray*}
 
The stochastic error of the complex Z$_2$ noise estimate results only
from the off-diagonal
entries of the inverse matrix (the same is true for Z$_n$ noise for
any n). However, other noises (such as gaussian)
have additional errors arising from diagonal entries. This is why the
Z$_2$ noise has minimum variance. For example, it has been demonstrated
on a $16^3 \times 24$ lattice with $\beta = 6.0$ and $\kappa = 0.148$ for
the Wilson action that the Z$_2$ noise standard deviation is smaller
than that of the gaussian noise by a factor of 1.54~\cite{dl94}.

Applying the complex Z$_2$ estimator to the expression for the
determinant ratio in Eq. (\ref{Pade_expand}), we find
\begin{eqnarray}
&&\sum_{k} b_k \{ \mbox {Tr}(M_1+c_k)^{-1}-
  \mbox {Tr}(M_2+c_k)^{-1} \} \nonumber\\
&&\approx \frac 1L \sum_{k}^K \sum_j^{L} b_k \eta^{j \dagger}
  [(M_1+c_k)^{-1} - (M_2+c_k)^{-1}]\eta^j  \nonumber\\
&&=  \frac 1L \sum_{j}^L \sum_{k=1}^{K} b_k \eta^{j \dagger} (\xi_1^{k,j}
  - \xi_2^{k,j} ),
\end{eqnarray}
 
where $\xi_{i}^{k,j}= (M_i+c_k {\bf I})^{-1} \eta^j$ are the solutions of
\begin{eqnarray}
(M_1 + c_k {\bf I} )\xi_1^{k,j} &=& \eta^j, ~~~~~~~\label{col_inv1}\\
(M_2 + c_k {\bf I} )\xi_2^{k,j} &=& \eta^j, ~~~~~~~ k,j=1, 2, \cdots~.
\label{col_inv2}
\end{eqnarray}
Since $M_i + c_k {\bf I}$ are shifted matrices with constant diagonal
matrix elements,
Eqs. (\ref{col_inv1}) and (\ref{col_inv2}) can be solved collectively
for all values of $c_k$
within one iterative process by several algorithms, including
the Quasi-Minimum Residual (QMR)
~\cite{fng95}, Multiple-Mass Minimum Residual (M$^3$ R)~\cite{ggl96},
and GMRES\cite{fg96}. We have adopted the M$^3$ R algorithm,
which has been shown to be about 2 times faster than the conjugate
gradient algorithm, and the overhead for the multiple $c_k$ is only
~8\%~\cite{ydl96}. The only price to pay is memory: for each $c_k$,
a vector of the solution needs to be stored.
Furthermore, one notices that $c_k > 0$ in Table 1. This improves the
conditioning of $(\mM + c_k \mI)$ since the eigenvalues of
$\mM$ have positive real parts.  Hence, we expect faster convergence for
column inversions for Eqs. (\ref{col_inv1}) and (\ref{col_inv2}).
 
In HMC, the difference between
the $\Tr \log \mM$ at the beginning and the end of the molecular
dynamics trajectory, i.e. $\Tr \log \mM_1 - \Tr \log \mM_2$ is
$\sim {\cal O}(1)$. Thus the standard deviation $\sigma$ encountered
in the estimation of $\Tr \log \mM_1 - \Tr \log \mM_2$ from the Z$_2$
noise
is of the same order as the estimated value itself. i.e.
$\sim {\cal O}(1)$. However, this is not good enough. In practice, we
find that one needs $ \sim 319,000$ noise vectors to reduce the stochastic
error to 0.2. In the next section,
we describe a method which significantly reduces the stochastic error.
 
\vspace{0.05in}
\section{Improved PZ Estimation with Unbiased Subtraction}
In order to reduce the variance of the estimate, we introduce a
suitably chosen set of traceless
$N \times N$ matrices $\mQ^{(p)}$, i.e. which satisfy
$\sum_{n=1}^{N} \mQ^{(p)}_{n,n} = 0,\, p = 1 \cdots P$.
The expected value and variance for  the modified estimator
$<\eta^\dagger(\mM^{-1} - \sum_{p=1}^P \gl_p \mQ^{(p)}) \eta>$
are given by
\begin{eqnarray}
E[<\eta^\dagger (\mM^{-1} - \sum_{p=1}^P \gl_p \mQ^{(p)})) \eta>]&=&
 \mbox {Tr }\mM^{-1}~, ~~~~~~~~~~~~~~~~~~~~~~~~~~~~~~~~~\\
\Delta_M(\gl) = \Var[<\eta^\dagger (\mM^{-1} -
                      \sum_{p=1}^P \gl_p \mQ^{(p)}) \eta>]
 &=& \frac 1L \sum_{m\neq n} |\mM^{-1}_{m,n}-
    \sum_{p=1}^P \gl_p \mQ^{(p)}_{m,n})|^2 ~,  \label{red_var}
\end{eqnarray}
for any values of the real parameters $\gl_p$. In other words,
introducing the matrices $\mQ^{(p)}$  into the estimator produces no
bias, but
may reduce the error bars if the $\mQ^{(p)}$ are chosen judiciously.
Further, $\gl_p$ may be varied at will to achieve a minimum
variance estimate:  this corresponds to a least-squares fit to the
function
$\eta^\dagger \mM^{-1} \eta$ sampled at points $\eta_j,~j=1 \cdots L$,
using the fitting
functions $\left\{1, \eta^\dagger \mQ^{(p)} \eta \right\},~
p=1 \cdots P$.
Making the definition $\mQ^{(0)} \equiv \mI$,
the usual least-square equations then yield
\begin{eqnarray*}
\mbox{ \boldmath $\gl$ } &=& \mC^{-1} \mbox{ \boldmath $\ga$},
     \mbox{~~~~~~where}\\
\mbox{ \boldmath $\gl$} &\equiv& (\gl_0,\gl_1,\cdots, \gl_P)^T,\\
\mC_{pq} &\equiv& \sum_{j=1}^L (\eta^{j \dagger}\mQ^{(p)} \eta^j)
                 (\eta^{j \dagger}\mQ^{(q)} \eta^j),\\
\ga_q &\equiv&
    \sum_{j=1}^L (\eta^{j \dagger}\mQ^{(q)} \eta^j)
    (\eta^{j \dagger}\mM^{-1} \eta^j),
\end{eqnarray*}
and the trace estimate is given by $N \cdot \gl_0$.
 
We now turn to the question of choosing suitable
traceless matrices $\mQ^{(p)}$  to use in the modified estimator.
One possibility for the Wilson fermion matrix $\mM = \mI - \kappa\mD$
is suggested by the hopping parameter --- $\kappa$ expansion of the
inverse matrix,
\begin{eqnarray}
(\mM + c_k \mI )^{-1} &=& \frac {1}{\mM + c_k \mI }
= \frac {1}{(1+c_k)(\mI - \frac {\kappa}{(1+c_k)} \mD) }\nonumber
   ~~~~~~~~~~~~~~~~~~~~~~~~\\
&=& \frac {\mI}{1+c_k} + \frac {\kappa}{(1+c_k)^2} \mD
+\frac {\kappa^2}{(1+c_k)^3} \mD^2
+\frac {\kappa^3}{(1+c_k)^4} \mD^3 + \cdots~.
\end{eqnarray}
This suggests choosing the matrices $\mQ^{(p)}$
from among those matrices in the hopping parameter expansion which
are traceless:
\begin{eqnarray*}
\mQ^{(1)} &=&\frac {\kappa}{(1+c_k)^2}\mD,~~~~~~~~~~~~~
                  ~~~~~~~~~~~~~~~~~~~~~\\
\mQ^{(2)} &=&\frac {\kappa^2}{(1+c_k)^3}\mD^2, \\
\mQ^{(3)} &=&\frac {\kappa^3}{(1+c_k)^4}\mD^3, \\
\mQ^{(4)} &=&\frac {\kappa^4}{(1+c_k)^5}(\mD^4-{\rm Tr}\mD^4), \\
\mQ^{(5)} &=&\frac {\kappa^5}{(1+c_k)^6}\mD^5, \\
\mQ^{(6)} &=&\frac {\kappa^6}{(1+c_k)^7}(\mD^6-{\rm Tr}\mD^6), \\
\mQ^{(2r+1)} &=& \frac {\kappa^{2r+1} }{(1+c_k)^{2r+2}}\mD^{2r+1},
~~~~~                          r=3,4,5, \cdots~.
\end{eqnarray*}
It may be verified that all of these matrices are traceless.
In principle, one can include all the even powers which entails the
explicit calculation of all the allowed loops in $Tr D^{2r}$. In this
manuscript we have only included $\mQ^{(4)}$, $\mQ^{(6)}$,
and $\mQ^{(2r+1)}$. Note that ${\Tr~}\mD^4$ in $\mQ^{(4)}$ can be
evaluated from
\begin{equation}
{\rm Tr~}\mD^4 = - 32 \sum_p {\rm Tr~}{\bf U}_p,
\end{equation}
where ${\Tr~}{\bf U}_p$ is the
plaquette and - 32 comes from the trace of the product of
$1 \pm \gamma_{\mu}$ in the the Wilson action. Similarly,
${\Tr~}\mD^6$ in $\mQ^{(6)}$ can be evaluated from 3 classes of
6-link loops,
\begin{equation}
\mbox{Tr}\{D^6\} = -128\sum_{L_1 \in R}U_{L_1}- 64\sum_{L_2 \in P}U_{L_2}
-64\sum_{L_3 \in C}U_{L_3},
\end{equation}
where $L_1$ stands for the sum over rectangles, $L_2$ over parallelograms, and
$L_3$ over chairs.

 We then set,
\begin{eqnarray}   \label{Q}
\mQ(\mbox{\boldmath $\gl$})=
\gl_1\mQ^{(1)}+\gl_2\mQ^{(2)}+\gl_3\mQ^{(3)}
+\gl_4\mQ^{(4)} +\gl_5\mQ^{(5)}+\gl_6\mQ^{(6)}+\cdots
+\gl_{2r+1}\mQ^{(2r+1)}+\cdots~,~~~~~~
\end{eqnarray}
and perform the variation process
to get an optimal choice of $\{~\gl_1,\gl_2,\cdots, \gl_{2r+1},
\cdots \}_{opt}$.
The additional computational cost incurred by the modified estimator is
$P$ additional matrix vector multiplications
per noise vector. Since $P$ is small ($\sim$ 9), this overhead is essentially
negligible compared to solving Eqs. (\ref{col_inv1}) and (\ref{col_inv2}).
 
~~~~~~~~~~~~~~~~~~~~~~~~~~~~~~~~~~\\
 
 
In actual practice, we generate $L$ complex Z$_2$ noise vectors,
and obtain basic PZ estimates using the M$^3$R matrix inversion algorithm.
The auxiliary data used in the improved PZ estimates may
be computed via a few matrix-vector multiplications.
\begin{itemize}
\item Unimproved estimates:~ $\{ O_1, O_2,\cdots, O_L\}$, ~with $O_j =
\sum_k b_k\eta^{j \dagger} \xi_{k,j}$,~~$j=1,2,\cdots,L$,
\item 1$^{st}$  auxiliary data set:~
$\{D^{(1)}_1, D^{(1)}_2,\cdots, D^{(1)}_L\}$, ~with $D^{(1)}_j =
\sum_k \frac {b_k\kappa^1 }{(1+c_k)^2 } (\eta^{j \dagger} \mD \eta^j)$,
\item 2$^{nd}$ auxiliary data set:~
$\{ D^{(2)}_1, D^{(2)}_2,\cdots, D^{(2)}_L\}$, ~with $D^{(2)}_j =
\sum_k \frac {b_k \kappa^2 }{(1+c_k)^3} (\eta^{j \dagger} \mD^{2}
\eta^j)$,
\item 3$^{rd}$ auxiliary data set:
$\{ D^{(3)}_1, D^{(3)}_2,\cdots, D^{(3)}_L\}$, with $D^{(3)}_j =
\sum_j \frac {b_k\kappa^3 }{(1+c_k)^4} (\eta^{j \dagger} \mD^{3}
\eta^j)$,
\item 4$^{th}$ auxiliary data set:
$\{ D^{(4)}_1, D^{(4)}_2,\cdots, D^{(4)}_L\}$, with $D^{(4)}_j =
\sum_k \frac {b_k\kappa^4 }{(1+c_k)^5} (\eta^{j \dagger} \mD^{4} \eta^j
                 -{\Tr}\mD^4)$,
\item 5$^{th}$ auxiliary data set:
$\{ D^{(5)}_1, D^{(5)}_2,\cdots, D^{(5)}_L\}$, ~with $D^{(5)}_j =
\sum_i \frac {b_i\kappa^5 }{(1+c_i)^6 }(\eta^{j \dagger} \mD^{5}
\eta^j)$,
\item 6$^{th}$ auxiliary data set:
$\{ D^{(6)}_1, D^{(6)}_2,\cdots, D^{(6)}_L\}$, with $D^{(6)}_j =
\sum_k \frac {b_k\kappa^6 }{(1+c_k)^7} (\eta^{j \dagger} \mD^{6} \eta^j
                 -{\Tr}\mD^6)$,
\item Higher odd-terms:
$\{ D^{(2r+1)}_1, D^{(2r+1)}_2,\cdots , D^{(2r+1)}_L \}$,
~with $D^{(2r+1)}_j =
\sum_i \frac {b_i\kappa^{2r+1} }{(1+c_i)^{2r+2} }(\eta^{j \dagger}
\mD^{2r+1} \eta^j)$,
\end{itemize}
Using these data, a least squares fit is performed to yield a set of
$\{\gl_0, \gl_1, \gl_2, \gl_3, \gl_4, \gl_5, \gl_6,
\gl_{(2r+1)} \}_{opt}$,
which minimizes the variance Eq. (\ref{red_var}) of the improved
estimator over the $\{ L \}$ noise vectors.
 
\section {Computations of determinants and determinant ratios}
 
Our numerical computations were carried out with the Wilson action
on the $8^3 \times 12$ ($N$ = 73728) lattice with $\beta =5.6$.
We use the HMC with pseudofermions to generate gauge configurations.
With a cold start, we obtain the fermion matrix $\mM_1$ after the
plaquette becomes stable. The trajectories are traced with $\tau = 0.01$
and 30 molecular dynamics steps using $\kappa = 0.150$.
$\mM_2$ is then obtained from $\mM_1$ by an accepted trajectory run.
Hence $\mM_1$ and $\mM_2$ differ by a continuum
perturbation, and $\log [\det \mM_1/\det \mM_2 ] \sim {\cal O}(1)$.
 
  We first calculate $\log \det\mM_1$ with different orders of Pad\'e
expansion around $z_0 = 0.1$ and $z_0 = 1.0$. We see from Table 2
that the 5th order Pad\'e does not give the same answer for two
different expansion points, suggesting that its accuracy is not
sufficient for the range of eigenvalues of $\mM_1$. Whereas, the 11th
order Pad\'e gives the same answer within errors.
Thus, we shall choose P[11,11]$(z)$ with $z_0=0.1$ to perform the
calculations from this point on.

Table 3 shows the optimal choice of parameters $\gl_i$, $i=1,5$
with different subtraction sets and various Z$_2$ noise lengths.
The fact that $\gl_i\sim 1.0$, $i=1,2,3,5$ and $\lambda_4 \sim 0$
gives further evidence that the fluctuations due to the
complex Z$_2$ noise are indeed introduced by the off-diagonal matrix
elements.

In Tables 4 and 5, we give the results of improved estimations for
$\Tr \log \mM_1$ and $\Tr \log \mM_2$ respectively.
We see that the variational technique
described above can reduce the data fluctuations by more than an order of
magnitude. For example, the unimproved error $\delta_0 = 5.54$ in
Table 4 for
400 Z$_2$ noises is reduced to $\delta_{11} = 0.15$ for the subtraction
which
includes up to the $\mQ^{11}$ matrix. This is 37 times smaller. Comparing
the central values in the last row (i.e. the $11^{th}$ order improved)
with that of unimproved estimate with 10,000 Z$_2$ noises, we see that
they are the same within errors. This verifies that the variational
subtraction scheme that we employed does not introduce biased
errors. The improved estimates of $\Tr \log \mM_1$ from 50
Z$_2$ noises with
errors $\delta_{r}$ from Table 4 are plotted in comparison with the central
value of the unimproved estimate from 10,000 noises in Fig. 8.

Our unimproved results have the similar size errors as those
obtained by the Chebyshev polynominal expansion of $\log \mM^{\dagger}\mM$
~\cite{sw97}, thus one can similarly improve its estimation with
the variational subtraction scheme introduced here.
\begin{table} \begin{center}
\caption{ Unimproved and improved PZ estimates for
log[det\mM$_1$]  with 100 complex Z$_2$ noise vectors. $\kappa = 0.150$.
 }
 
\begin{tabular}{c c c c c c } \hline\hline
$P[K,K](z)$ &  $K=$  &  5  &   7   &  9   &  11     \\ \hline
$z_0 = 0.1$ & Original: & 473(10) & 774(10) & 796(10) & 798(10) \\
~ & Improved: & 487.25(62) & 788.17(62) & 810.83(62) & 812.33(62) \\ \hline
$z_0 = 1.0$ &Original: & 798(10) & 798(10) & 798(10) & 799(10)\\
~ & Improved: & 812.60(62) & 812.37(62) & 812.36(62) & 812.37(62)\\
\hline\hline
\end{tabular}
\end{center} \end{table}

\newpage
\clearpage
{\small
\begin{table} \begin{center}
\caption{The variational parameters $\{ \gl \}_{opt}$ in Eq. (\ref{Q})
for different subtraction data set listed above and various Z$_2$ noise
length $L$ for \mM$_1$.}
\begin{tabular}{cc|c c c c c } \hline\hline
Improved & Z$_2$ length & $\lambda_1$ & $\lambda_2$ & $\lambda_3$ &
$\lambda_4$ & $\lambda_5$
\\ \hline
$1^{st}$: & $L$=200   & 0.960  &  ---   &  ---  &  ---  & --- \\
$L$=600   & 0.968  &  ---   &  ---   &  ---  & --- \\  \hline
$2^{nd}$: & $L$=200   & 0.960  & 0.919  &  ---  &  ---  & --- \\
& $L$=600   & 0.972  & 0.982  &  ---   &  ---  & ---  \\  \hline
$3^{rd}$: & $L$=200   & 0.968  & 0.942  & 0.908 &  ---  & --- \\
& $L$=600   & 0.972  & 0.969  & 0.919  &  ---  & ---  \\  \hline
$4^{th}$: & $L$=200   & 0.973  & 0.921  & 0.903 & 0.092 & --- \\
& $L$=600   & 0.976  & 0.924  & 0.967  & 0.089 & ---  \\  \hline
$5^{th}$: & $L$=200   & 0.990  & 0.926  & 0.908 & 0.089 & 1.27 \\
& $L$=600   & 0.992  & 0.924  & 0.910  & 0.087 & 1.26 \\ \hline \hline
$11^{th}$: &$L=50$  & 0.998  & 0.913  & 0.968 & 0.088 & 1.06 \\
&$L=100$& 0.992  & 0.890  & 0.953 & 0.086 & 1.48 \\
&$L=150$& 0.995  & 0.886  & 0.939 & 0.086 & 1.18 \\
&$L=200$& 0.999  & 0.923  & 0.951 & 0.086 & 1.09 \\
&$L=300$& 0.997  & 0.924  & 0.934 & 0.087 & 1.06 \\
&$L=400$& 0.998  & 0.925  & 0.959 & 0.088 & 1.07 \\
&$L=600$& 1.001  & 0.923  & 0.964 & 0.085 & 1.00
\\ \hline \hline
\end{tabular}
\caption{ Central values for improved stochastic estimation of
log[det {\bf M}$_1$] and $r$th--order improved Jackknife errors
$\delta_r$ are given for different numbers of Z$_2$ noise vectors.
$\kappa$ is 0.150 in this case.
}
\begin{tabular}{ c c c c c c c c c c} \hline\hline
 \# Z$_2$    & 50 & 100 & 200 & 400 & 600 & 800 & 1000 & 3000 & 10000 \\ \hline
 0$^{th}$    &  802.98   &  797.98   &  810.97   &  816.78   &  815.89
             &  813.10   &  816.53   &  813.15   &  812.81    \\
 $\delta_0$  & $\pm$14.0 & $\pm$9.81 & $\pm$7.95 & $\pm$5.54 & $\pm$4.47
	     & $\pm$3.83 & $\pm$3.41 & $\pm$1.97 & $\pm$1.08  \\ \hline
  1$^{st}$   &  807.89   &  811.21   &  814.13   &  815.11   &  814.01
             &  814.62   &  814.97   & --- & ---  \\
 $\delta_1$  & $\pm$4.65 & $\pm$3.28 & $\pm$2.48 & $\pm$1.84 & $\pm$1.50
             & $\pm$1.29 & $\pm$1.12 &  -  & -    \\ \hline
  2$^{nd}$   &  813.03   &  812.50   &  811.99   &  812.86   &  811.87
             &  812.89   &  813.04 & ---  & ---   \\
 $\delta_2$  & $\pm$2.46 & $\pm$1.65 & $\pm$1.34 & $\pm$1.01 & $\pm$0.83
             & $\pm$0.72 & $\pm$0.64 & -   & -    \\ \hline
  3$^{rd}$   &  812.07   &  812.01   &  811.79   &  812.44   &  812.18
             &  812.99   &  813.03 & ---  & ---   \\
 $\delta_3$  & $\pm$1.88 & $\pm$1.31 & $\pm$1.01 & $\pm$0.74 & $\pm$0.58
             & $\pm$0.51 & $\pm$0.44 & -  & -    \\ \hline
  4$^{th}$   &  812.28   &  812.52   &  812.57   &  812.86   &  812.85
             &  813.25   &  813.40 & ---   & ---  \\
 $\delta_4$  & $\pm$1.20 & $\pm$0.94 & $\pm$0.68 & $\pm$0.48 & $\pm$0.39
             & $\pm$0.35 & $\pm$0.30 & -  & -    \\ \hline
  5$^{th}$   &  813.50   &  813.07   &  813.36   &  813.70   &  813.47
             &  813.54   &  813.50 & ---  & ---  \\
 $\delta_5$  & $\pm$0.82 & $\pm$0.62 & $\pm$0.47 & $\pm$0.34 & $\pm$0.29
             & $\pm$0.25 & $\pm$0.22 & -  & -    \\ \hline
  6$^{ts}$   &  813.54   &  813.23   &  813.22   &  813.28   &  813.20
	     &  813.37   &  813.26 & ---  & ---  \\
 $\delta_6$  & $\pm$0.67 & $\pm$0.49 & $\pm$0.35 & $\pm$0.25 & $\pm$0.21
             & $\pm$0.18 & $\pm$0.16 & -  & -    \\ \hline
  7$^{ts}$   &  814.18   &  813.74   &  813.44   &  813.42   &  813.39
             &    ---    &    ---  & ---  & ---  \\
 $\delta_7$  & $\pm$0.44 & $\pm$0.36 & $\pm$0.26 & $\pm$0.19 & $\pm$0.16
             &     -     &     -     & -  & -    \\ \hline
  9$^{th}$   &  813.77   &  813.62   &  813.49   &  813.40   &  813.43
             &    ---    &    ---  & ---  & ---  \\
 $\delta_9$  & $\pm$0.40 & $\pm$0.30 & $\pm$0.22 & $\pm$0.16 & $\pm$0.14
	     &     -     &     -     & -  & -    \\ \hline
 11$^{th}$   &  813.54   &  813.41   &  813.45   &  813.44   &  813.43
             &    ---    &    ---  & ---  & ---  \\
$\delta_{11}$& $\pm$0.38 & $\pm$0.27 & $\pm$0.21 & $\pm$0.15 & $\pm$0.13
             &     -     &     -     & - & -    \\ \hline\hline
\end{tabular}
\end{center} \end{table}
}
\newpage
\clearpage
\begin{table} \begin{center}
\caption{ The same as in Table 4 for log[det {\bf M}$_2$].
}
\begin{tabular}{ c c c c c c c c c c} \hline\hline
\# Z$_2$=  & 50 & 100 & 200 & 400 & 600 & 800 & 1000 & 3000 & 10000  \\ \hline
 0$^{th}$    &  788.96   &  793.87   &  809.08   &  813.03   &  813.06
      &  811.14   &  815.29   &  811.48   &  809.54    \\
 $\delta_0$  & $\pm$15.8 & $\pm$10.7 & $\pm$8.27 & $\pm$5.89 & $\pm$4.66
	    & $\pm$3.94 & $\pm$3.46 & $\pm$1.99 & $\pm$1.09  \\ \hline
 1$^{st}$   &  808.94   &  814.00   &  811.94   &  811.85   &  811.73
	   &  811.77   &  812.22   & --- & ---  \\
 $\delta_1$  & $\pm$5.00 & $\pm$3.69 & $\pm$2.69 & $\pm$1.91 & $\pm$1.53
 & $\pm$1.39 & $\pm$1.25 &  -  & -    \\ \hline
 2$^{nd}$   &  810.75   &  811.11   &  810.29   &  810.21   &  810.15
&  810.43   &  811.05 & ---  & ---   \\
 $\delta_2$  & $\pm$2.67 & $\pm$1.94 & $\pm$1.40 & $\pm$1.03 & $\pm$0.84
  & $\pm$0.73 & $\pm$0.66 & -   & -    \\ \hline
3$^{rd}$   &  806.65   &  808.13   &  809.28   &  809.58   &  810.06
    &  810.56   &  810.52 & ---  & ---   \\
$\delta_3$  & $\pm$1.79 & $\pm$1.36 & $\pm$1.01 & $\pm$0.69 & $\pm$0.56
 & $\pm$0.48 & $\pm$0.43 & -  & -    \\ \hline
4$^{th}$   &  807.80   &  808.89   &  810.02   &  809.76   &  810.06
 &  810.32   &  810.40 & ---   & ---  \\
$\delta_4$  & $\pm$1.17 & $\pm$0.97 & $\pm$0.69 & $\pm$0.49 & $\pm$0.39
& $\pm$0.33 & $\pm$0.30 & -  & -    \\ \hline
 5$^{th}$   &  809.92   &  809.80   &  810.55   &  810.68   &  810.82
   &  810.85   &  810.79 & ---  & ---  \\
$\delta_5$  & $\pm$0.89 & $\pm$0.69 & $\pm$0.52 & $\pm$0.37 & $\pm$0.29
  & $\pm$0.25 & $\pm$0.22 & -  & -    \\ \hline
 7$^{ts}$   &  810.47   &  809.93   &  810.65   &  810.58   &  810.74
  &    ---    &    ---  & ---  & ---  \\
  $\delta_7$  & $\pm$0.75 & $\pm$0.62 & $\pm$0.46 & $\pm$0.32 & $\pm$0.25
  &     -     &     -     & -  & -    \\ \hline
9$^{th}$   &  809.04   &  809.90   &  810.70   &  810.71   &  810.80
 &    ---    &    ---  & ---  & ---  \\
$\delta_9$  & $\pm$0.66 & $\pm$0.59 & $\pm$0.43 & $\pm$0.29 & $\pm$0.23
 &     -     &     -     & -  & -    \\ \hline
11$^{th}$   &  810.02   &  809.73   &  810.65   &  810.71   &  810.81
 &    ---    &    ---  & ---  & ---  \\
$\delta_{11}$& $\pm$0.65 & $\pm$0.56 & $\pm$0.42 & $\pm$0.29 & $\pm$0.23
  &     -     &     -     & - & -    \\ \hline\hline
\end{tabular}
\end{center} \end{table}
\newpage
\clearpage
\begin{figure}[th]
\setlength\epsfxsize{120mm}
\epsfbox{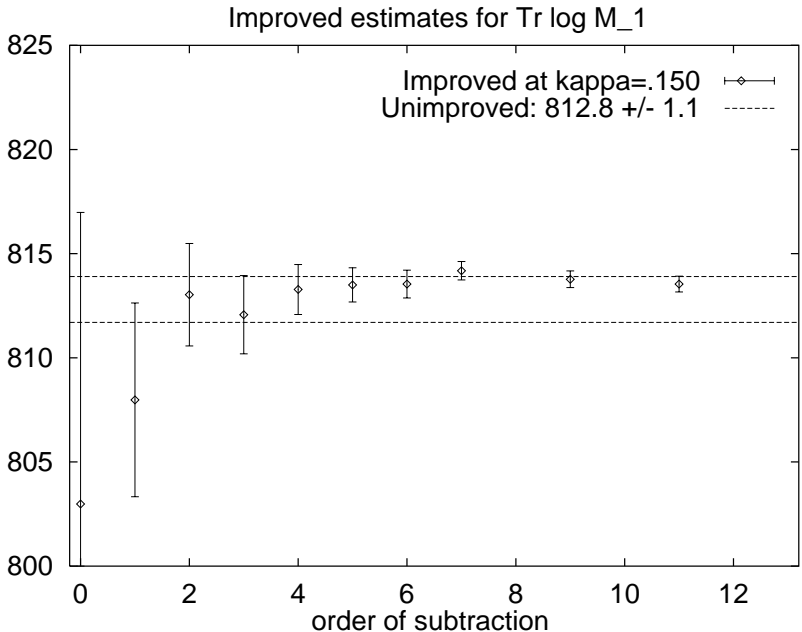}
\caption{The improved PZ estimate of $\Tr \log \mM_1$ with 50 noises
as a function
of the order of subtraction and compared to that of unimproved estimate
with 10,000 noises. The dashed lines are drawn with a distance of
1 $\sigma$ away from the central value of the unimproved estimate.}
\end{figure}
 
Results for $\Tr[\log \mM_1 -\log \mM_2 ]$ are shown in Table 6.
We see that again the errors are reduced by a factor $\sim 34$.
$\delta_{11}$ for 50 Z$_2$ noise vectors is even smaller than the unimproved
error $\delta_0$ with L = 10,000. To achieve the same level of accuracy
for the unimproved
estimation, it would require $\sim 65,955$ noise vectors. This is
1,319 times more than the 50 noise case which employs subtraction.
Again, to show that the subtraction does not introduce biased errors, we
plot in Fig. 9 the improved PZ estimates of $\Tr[\log \mM_1 -\log \mM_2 ]$
with errors from 50 noise vectors as a function of the order of subtraction
and verify that they agree with that of the unimproved estimate with 10,000
noises.
 
As for the quark mass dependence, we expect that the error will go
up as the quark mass becomes smaller. Indeed, we show in Table 7 the
results of $\Tr[\log \mM_1 -\log \mM_2 ]$ for
$\kappa = 0.154$ that the errors are larger.
The PZ estimates and their errors in Table 7 for $\kappa = 0.154$ are
similarly plotted as a function of the order of subtraction in Fig. 10.

We have also generated a sequence of configurations through HMC updating
with \mbox{pseudofermions}. In Table 8 we list the change of the gauge,
the pseudofermion, and the kinetic energy parts of the
action from 10 molecular dynamics trajectories. The total
change in energy $\Delta H$ is $\sim O(1)$. Also listed are the change of
$\Tr \log \mM$,
i.e. $\Delta (\Tr \log \mM) = \Tr \log \mM_1 - \Tr \log \mM_2$.
It is somewhat surprising to see that the absolute values of
$\Delta (\Tr \log \mM)$ are an order of magnitude smaller that those
of $\Delta U_{pseudo}$ (the pseudofermion part of the action), and their
signs can be different. This
may be related to the observation that it takes very long to decorrelate
the global topological charge in HMC with pseudofermions~\cite{abe96}.
This will be investigated further in the future.
\newpage
\clearpage
\begin{figure}[h]
\setlength\epsfxsize{120mm}
\epsfbox{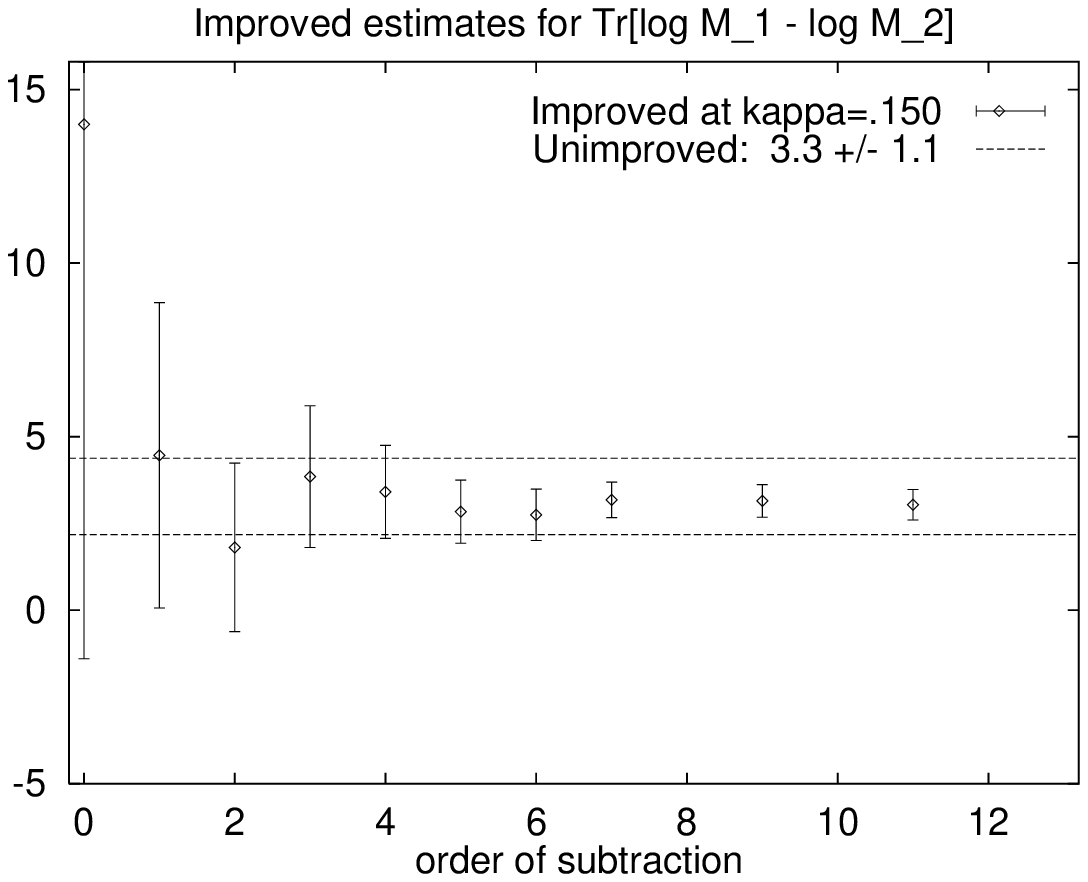}
\caption{The same as in Fig. 8 for $\Tr[ \log \mM_1 - \log \mM_2]$
with $\kappa=0.150$.}
\vspace{0.2in}
\epsfbox{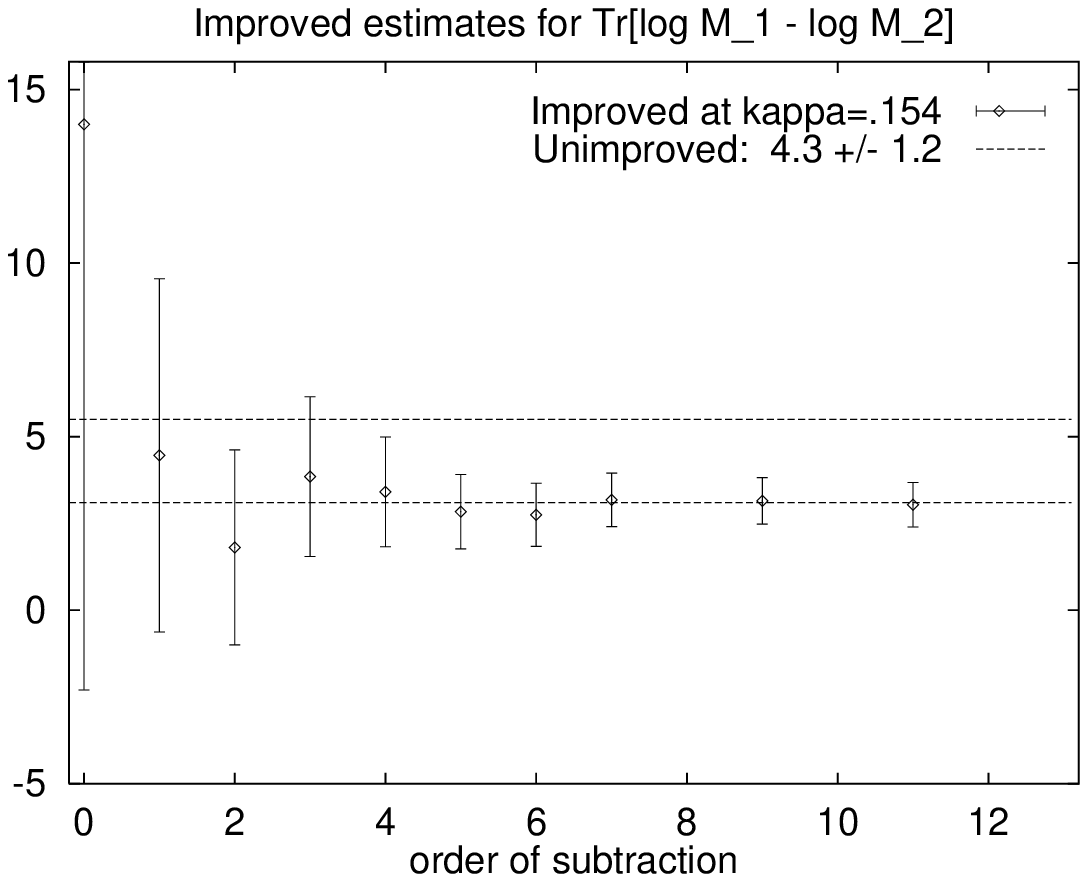}
\caption{The same as in Fig. 9 for $\Tr[ \log \mM_1 - \log \mM_2]$
with $\kappa = 0.154$.}
\end{figure}
\newpage
\clearpage
\begin{table} \begin{center}
\caption{ The same as in Table 4 for log[det {\bf M}$_1$ /
det {\bf M}$_2$]. }
 
\begin{tabular}{ c c c c c c c c c c c} \hline\hline
$L$  & 50 & 100 & 200 & 400 & 600 &  800 & 1000 & 3000 & 10000 \\ \hline
 0$^{th}$    &    14.0   &   4.12    &   1.90    &   3.75    &   2.84
             &    1.96   &   1.24    &   1.67    &   3.28    &  \\
 $\delta_0$  & $\pm$15.4 & $\pm$11.5 & $\pm$8.20 & $\pm$5.59 & $\pm$4.68
             & $\pm$4.01 & $\pm$3.54 & $\pm$2.06 & $\pm$1.13 &  \\ \hline
  1$^{st}$   &    4.46   &    2.39   &    2.30   &    3.23   &  2.24
             &    2.85   &    2.75   &    3.16   & ---  & \\
 $\delta_1$  & $\pm$4.40 & $\pm$3.56 & $\pm$2.75 & $\pm$1.99 & $\pm$1.67
             & $\pm$1.45 & $\pm$1.31 & $\pm$0.75 & - &  \\ \hline
  2$^{nd}$   &    1.81   &    1.49   &    1.72   &    2.62   &  1.75
             &    2.45   &    1.99   &    3.05   & ---  &  \\
 $\delta_2$  & $\pm$2.43 & $\pm$1.98 & $\pm$1.49 & $\pm$1.11 & $\pm$0.95
             & $\pm$0.82 & $\pm$0.73 & $\pm$0.42 & -  &  \\ \hline
  3$^{rd}$   &    3.85   &    2.72   &    2.45   &    2.78   &  2.16
             &    2.43   &    2.48   &    3.04   & ---  &  \\
 $\delta_3$  & $\pm$2.04 & $\pm$1.45 & $\pm$1.05 & $\pm$0.76 & $\pm$0.63
             & $\pm$0.54 & $\pm$0.48 & $\pm$0.27 & -   &  \\ \hline
  4$^{th}$   &    3.41   &    2.77   &    2.51   &    3.04   &  2.81
             &    2.92   &    2.99   & ---  & ---   & \\
 $\delta_4$  & $\pm$1.34 & $\pm$0.97 & $\pm$0.70 & $\pm$0.51 & $\pm$0.42
             & $\pm$0.36 & $\pm$0.33 & -  & -   & \\ \hline
  5$^{th}$   &    2.84   &    2.84   &    2.75   &    2.95   &  2.62
             &    2.68   &    2.70 & ---  & ---   & \\
 $\delta_5$  & $\pm$0.91 & $\pm$0.65 & $\pm$0.51 & $\pm$0.38 & $\pm$0.32
             & $\pm$0.27 & $\pm$0.24 & -  & -   & \\ \hline
  6$^{th}$   &    2.75   &    2.80   &    2.65   &    2.65   &  2.49
             &    2.54   &    2.51 & ---  & ---   & \\
 $\delta_6$  & $\pm$0.74 & $\pm$0.52 & $\pm$0.39 & $\pm$0.29 & $\pm$0.24
             & $\pm$0.20 & $\pm$0.18 & -  & -   & \\ \hline
  7$^{ts}$   &    3.18   &    3.27   &    2.77   &    2.90   &  2.77
             &     ---   &    ---    & ---  & ---   & \\
 $\delta_7$  & $\pm$0.51 & $\pm$0.35 & $\pm$0.27 & $\pm$0.21 & $\pm$0.18
             &     -     &     -     & -  & -   & \\ \hline
  9$^{th}$   &    3.15   &    3.39   &    2.78   &    2.81   &  2.76
             &     ---   &    ---  & ---  & ---   & \\
 $\delta_9$  & $\pm$0.47 & $\pm$0.32 & $\pm$0.25 & $\pm$0.19 & $\pm$0.16
             &     -     &     -     & -  & -   & \\ \hline
 11$^{th}$   &    3.04   &    3.33   &    2.76   &    2.83   &  2.72
             &    ---    &    ---    & ---   & ---  &  \\
$\delta_{11}$& $\pm$0.44 & $\pm$0.29 & $\pm$0.23 & $\pm$0.17 & $\pm$0.14
             &    -      &     -     &  -   & -  &  \\ \hline\hline
\end{tabular}
\end{center} \end{table}
\begin{table} \begin{center}
\caption{Same as Table 6 for the log[det {\bf M}$_1$/det {\bf M}$_2$]
at $\kappa=.154$ instead of $\kappa=.150$.
}
 
\begin{tabular}{ c c c c c c c c c c c} \hline\hline
 Z$_2$ $^\#$ & 50 & 100 & 200 & 400 & 600 & 800  & 1000 & 3000 & 10000 \\
 \hline
 0$^{th}$    &    12.6   &   2.66    &   2.32    &   4.86    &   3.84    &
                  2.68   &   1.92    &   3.35    &   4.31    &  \\
 $\delta_0$  & $\pm$16.3 & $\pm$12.0 & $\pm$8.53 & $\pm$5.83 & $\pm$4.86 &
               $\pm$4.18 & $\pm$3.68 & $\pm$2.15 & $\pm$1.19 &  \\ \hline
  1$^{st}$   &   -0.45   &   -1.60   &    3.56   &    4.67   &  3.27   &
                  4.11   &    4.04   & --- & ---  & \\
 $\delta_1$  & $\pm$5.09 & $\pm$3.88 & $\pm$2.96 & $\pm$2.16 & $\pm$1.79 &
               $\pm$1.55 & $\pm$1.39 & -  & -   &  \\ \hline
  2$^{st}$   &    3.54   &    3.21   &    3.17   &    4.14   &  2.88   &
                  3.70   &    3.32   & --- & ---  & \\
 $\delta_2$  & $\pm$2.81 & $\pm$2.27 & $\pm$1.66 & $\pm$1.26 & $\pm$1.08 &
               $\pm$0.93 & $\pm$0.69 & -  & -   &  \\ \hline
  3$^{rd}$   &    6.27   &    5.07   &    3.90   &    4.32   &  3.58   &
                  3.84   &    3.94   & --- & ---  &  \\
 $\delta_3$  & $\pm$2.30 & $\pm$1.65 & $\pm$1.20 & $\pm$0.89 & $\pm$0.74 &
               $\pm$0.64 & $\pm$0.57 & -  & -   &  \\ \hline
  4$^{th}$   &    5.49   &    5.22   &    4.28   &    4.73   &  4.46   &
                  4.49   &    4.49 & ---  & ---   & \\
 $\delta_4$  & $\pm$1.58 & $\pm$1.13 & $\pm$0.54 & $\pm$0.64 & $\pm$0.52 &
               $\pm$0.45 & $\pm$0.41 & -  & -   & \\ \hline
  5$^{th}$   &    4.55   &    4.72   &    4.30   &    4.35   &  4.01   &
                  4.10   &    4.08 & ---  & ---   & \\
 $\delta_5$  & $\pm$1.07 & $\pm$0.77 & $\pm$0.63 & $\pm$0.49 & $\pm$0.41 &
               $\pm$0.35 & $\pm$0.32 & -  & -   & \\ \hline
  6$^{ts}$   &    4.51   &    4.72   &    4.12   &    4.03   &  3.83   &
                  3.91   &    3.88   & ---  & ---   & \\
 $\delta_6$  & $\pm$0.91 & $\pm$0.68 & $\pm$0.54 & $\pm$0.42 & $\pm$0.34 &
               $\pm$0.29 & $\pm$0.26 & -  & -   & \\ \hline
  7$^{ts}$   &    5.13   &    5.52   &    4.35   &    4.43   &  4.24   &
                  4.16   &    4.02   & ---  & ---   & \\
 $\delta_7$  & $\pm$0.77 & $\pm$0.55 & $\pm$0.44 & $\pm$0.35 & $\pm$0.28 &
               $\pm$0.24 & $\pm$0.22 & -  & -   & \\ \hline
  9$^{th}$   &    4.98   &    5.60   &    4.33   &    4.27   &  4.21   &
                  4.16   &    4.08   & ---  & ---   & \\
 $\delta_9$  & $\pm$0.67 & $\pm$0.52 & $\pm$0.43 & $\pm$0.32 & $\pm$0.26 &
               $\pm$0.22 & $\pm$0.20 & -  & -   & \\ \hline
 11$^{th}$   &    4.80   &    5.60   &    4.32   &    4.31   &  4.19   &
                  4.21   &    4.13   & ---   & ---  &  \\
$\delta_{11}$& $\pm$0.64 & $\pm$0.49 & $\pm$0.42 & $\pm$0.31 & $\pm$0.25 &
               $\pm$0.21 & $\pm$0.19 &  -   & -  &  \\ \hline\hline
\end{tabular}
\end{center} \end{table}
\begin{table} \begin{center}
\caption{A breakdown of the energy change $\Delta H$ in 10
molecular dynamics trajectories in terms of the change in gauge
action ($\Delta U_{plaq}$), pseudofermion action ($\Delta U_{pseudo}$),
and kinetic energy ($\Delta \pi^2$). Also listed are the
estimates of $\Delta (Tr \log \mM) = \Tr \log \mM_1 -
\Tr \log \mM_2$ using 600 Z$_2$ noises with subtraction. The $\kappa$
is 0.150 in this case.}
 
\begin{tabular}{c c c c } \hline\hline
Group & ~ & $\Delta H_(old - new)$ & $\Delta (Tr \log \mM)$\\ \hline
Pair 1: & $\Delta U_{plaq}$ &-188.238 &    \\
$M_1, M_2$ & $\Delta U_{pseudo}$ &  15.636 &-0.21(.24) \\
~& $\Delta \pi^2$  &  172.514 &           \\
\hline
Pair 2: & $\Delta U_{plaq}$ & 437.556 &    \\
$M_2, M_3$ & $\Delta U_{pseudo}$  &  15.518 &-3.58(.26) \\
~& $\Delta \pi^2$ & -453.023 &           \\
\hline
Pair 3: & $\Delta U_{plaq}$  &-120.857 &    \\
$M_3, M_4$ & $\Delta U_{pseudo}$  &  -6.505 & 0.56(.25) \\
~& $\Delta \pi^2$ & 127.331 &           \\
\hline
Pair 4: & $\Delta U_{plaq}$ & -40.862 &    \\
$M_4, M_5$ & $\Delta U_{pseudo}$  & -28.189 & 2.03(.24) \\
~& $\Delta \pi^2$ & 69.085 &           \\
\hline
Pair 5: & $\Delta U_{plaq}$ &-110.674 &    \\
$M_5, M_6$ & $\Delta U_{pseudo}$  & -70.823 &-0.06(.24) \\
~& $\Delta \pi^2$  & 181.516 &           \\
\hline
Pair 6: & $\Delta U_{plaq}$ & 241.814 &    \\
$M_6, M_7$ & $\Delta U_{pseudo}$  &  67.498 &-3.69(.25) \\
~& $\Delta \pi^2$  & -309.213 &           \\
\hline
Pair 7: & $\Delta U_{plaq}$  &  82.339 &    \\
$M_7, M_8$ & $\Delta U_{pseudo}$   & -70.015 &-0.98(.24) \\
~& $\Delta \pi^2$  & -12.315 &           \\
\hline
Pair 8: & $\Delta U_{plaq}$  &-692.873 &    \\
$M_8, M_9$ & $\Delta U_{pseudo}$  & -27.643 & 6.51(.24) \\
~& $\Delta \pi^2$  & 720.243 &           \\
\hline
Pair 9: & $\Delta U_{plaq}$  & 260.435 &    \\
$M_9, M_{10}$ & $\Delta U_{pseudo}$   &   2.186 &-1.14(.25) \\
~& $\Delta \pi^2$  & -262.522 &           \\
\hline
Pair 10: &$\Delta U_{plaq}$  &-613.121 &    \\
$M_{10},M_{11}$ & $\Delta U_{pseudo}$   & 110.997 & 6.40(.24) \\
~& $\Delta \pi^2$  & 501.946 &           \\
\hline
Ex. Pair: & $M_1, M_{11}$  & ~ &   5.85(.30) \\
\hline \hline
\end{tabular} \end{center}
\end{table}
\newpage
\clearpage
\section{Application to global density of states for Hermitian
     Hamiltonian systems}
The density of states $\gr(z)$ for a Hamiltonian system with
Hamiltonian matrix $\mH$ is
\begin{equation}
     \gr(z) = \frac 1N \sum_{n=1}^{N} \gd (z - \gl_n),
\end{equation}
where $\left\{ \gl_n \right\}$ are the eigenvalues of $\mH$.

In reference \cite{wbd80}, $\gr(z)$ for real $z$ is calculated for
Hermitian $\mH$ as follows:
\begin{equation}
     \gr(z) = \frac{1}{\pi}\lim_{\gep \rightarrow 0^+}
   \mbox{Im} \Tr (\mH - (z+\gi \gep)\mI)^{-1}.
\label{Green_DOS}
\end{equation}
Choosing a small $\gep$  in Equation (\ref{Green_DOS}) yields a
smoothed version of $\gr(z)$.

We have shown above that the trace in
Eq. (\ref{Green_DOS}) may be estimated for several different
values of $z$ simultaneously using complex Z$_2$ noise and the M$^3$R
algorithm.  Thus, we may estimate the global density of states for a
Hermitian matrix $\mH$ at essentially the same computational cost
(modulo additional memory) as estimating the local density of
states at a single point.

\section{Conclusion and summary of advantages of the PZ algorithm}
The PZ method takes advantage of proven, effective numerical
approximation techniques. The advantages of the PZ method are summarized as
follows:

\begin{itemize}
\item
Pad\'e approximation uses rational functions,
which are known to be very efficient in the uniform approximation of
analytic functions. In finding determinant ratios, the Pad\'{e}
approximation to
the logarithm only needs to be accurate on the region in the complex
plane where the spectra of ${\bf M}_{1}$  and ${\bf M}_{2}$ differ.

\item
The complex $Z_{2}$ random vectors have been shown to be superior to the
gaussian~\cite{dl94,dll95,Eick96} noise in computing traces of
inverse matrices.

\item
The PZ method also takes advantage of the recently developed M$^3$R
algorithm to calculate all terms in the Pad\'e
expansion in Eq. (\ref{Pade_expand}) in essentially the same computational
time required to calculate a single term, albeit with additional memory
(one length $N$ vector for each additional term).
Hence, a higher order Pad\'{e} expansion requires more memory,
but essentially the same computation time (apart from the matrix
conditioning effects to be mentioned below).

\item
The entire method can be
applied to non-Hermitian matrices:  so determinants of non-Hermitian
matrices may also be found directly, without recourse to the Hermitian
matrix ${\mM}^{\dagger}{\mM}$. Negative and complex determinants can
also be calculated in principle.

\item
The  $c_{j}$'s in the Pad\'e expansion in Eq. (\ref{Pade_apprx})
turn out to be real and positive,
which improves the conditioning of the matrices ${\bf M}+c_{j}{\bf I}$ and
hence expedites the column inversions in
Eqs. (\ref{col_inv1}, \ref{col_inv2}).
However, this effect diminishes for higher order Pad\'e approximations,
because the minimum $c_j$ decreases as the order increases.

\item
The PZ method also holds promise of being useful in the case where
different quark flavors are present, in which case it is necessary to
compute multiple determinant ratios for matrices with different but constant
diagonal terms.
Using the M$^3$R algorithm, this takes essentially the same computation time
as a single determinant ratio.

\item
The unbiased variational subtraction scheme works quite well in reducing
the stochastic error from the Z$_2$ noise. The principle is general enough
to be applied to other cases with stochastic estimates.

\end{itemize}

 In conclusion, we have demonstrated the efficiency of
the Pad\'e - Z$_2$ algorithm in estimating determinants and determinant
ratios to high accuracy for
lattice QCD. It is certainly applicable to other systems with large
sparse matrices. For example, we have been able to reduce the error of the
determinant
ratio from 559\% to 17\% with the unbiased subtraction scheme and a
relatively small ($\sim$ 400) number of the noise vectors (see Table 6).
It is rather encouraging as far as the the feasibility
of using this algorithm to simulate dynamical
fermions in full QCD is concerned. We shall pursue this in the future.

\vskip 1.0cm
\noindent
{\it Acknowledgments} This work is partially supported by U.S. DOE grant
No. DE-FG05-84ER40154. The authors would like to thank Z. Bai, P. deForcrand,
A. Frommer, G. Goulob, A. Kennedy, and D. Weingarten for helpful discussions.
They would also like to thank \mbox{C. Liu} for lending us his HMC code and
W.C. Kuo for providing results on the 6-link loops.

\end{document}